\documentclass{aa}
\usepackage{natbib}
\usepackage{graphicx}
\usepackage{txfonts}
\bibpunct{(}{)}{;}{a}{}{,}
\newcommand{\Frac}[2]{\frac{\displaystyle\strut #1}{\displaystyle\strut #2} }

\begin{document}

\title{The empirical Earth rotation model from VLBI observations}
\author{L. Petrov\thanks{On leave from NVI, Inc.}}
\institute{National Astronomical Observatory of Japan, Mizusawa,
           2-12, Hoshigaoka-cho, Mizusawa-ku, Oshu-shi, Iwate-ken, 
           023-0861, Japan 
           \email{leonid.petrov@lpetrov.net} }

\date{{Received 2006.02.26/Accepted 2006.12.13}, \enskip A \& A, \enskip
 {\bf 467}, \enskip 359--369(2007) \enskip DOI: 10.1051/000406361:20065091}
\abstract
{}
{An alternative to the traditional method for modeling kinematics of the 
Earth's rotation is proposed. The purpose of developing the new approach 
is to provide a self-consistent and simple description of the Earth's 
rotation in a way that can be estimated directly from observations 
without using intermediate quantities.}
{Instead of estimating the time series of pole coordinates, the UT1--TAI 
angles, their rates, and the daily offsets of nutation, it is proposed 
to estimate coefficients of the expansion of a small perturbational rotation 
vector into basis functions. The resulting transformation from the terrestrial 
coordinate system to the celestial coordinate system is formulated as 
a product of an a priori matrix of a finite rotation and an empirical 
vector of a residual perturbational rotation. In the framework of this 
approach, the specific choice of the a~priori matrix is irrelevant, provided 
the angles of the residual rotation are small enough to neglect their 
squares. The coefficients of the expansion into the B-spline and 
Fourier bases, together with estimates of other nuisance parameters, are 
evaluated directly from observations of time delay or time range in a 
single least square solution.}
{This approach was successfully implemented in a computer program for 
processing VLBI observations. The dataset from 1984 through 2006 was 
analyzed. The new procedure adequately represents the Earth's rotation, 
including slowly varying changes in UT1--TAI and polar motion, the forced 
nutations, the free core nutation, and the high frequency variations 
of polar motion and UT1.}
{}


\maketitle

\section{Introduction}

  As we learn from an elementary physics course in high school, the rotation of
a rigid body can be represented by three angles. Euler angles are usually
selected as parameters, although it is not the only choice. We can say that
rotation of an arbitrary body is defined if a functional dependence of Euler
angles on time is known. 

  However, when we are dealing with the Earth, the traditional way of 
representing the kinematics of the Earth's rotation is not as simple. 
In modern textbooks that follow the formalism of Newcomb-Andoyer, for instance, 
\citet{r:GreenBook}, the Earth's rotation is represented as a product 
of 9 matrices that depend on many parameters. Analytical expressions are 
provided for some of them, but other parameters are supposed to be determined 
from observations. These parameters are defined though intermediate quantities. 
These intermediate quantities like the Earth's rotation axis, the celestial 
ephemeris pole, the true equinox of date, the non-rotating origin and others, 
are not objects that can be observed, but idealistic concepts that do not 
have a clear, intuitive interpretation. 

  According to the traditional approach, \citep[e.g.][]{r:iers2003}, 
in order to get the Earth's orientation at any given moment, one should 
first compute the values of intermediate parameters that have more than 
a thousand terms, then interpolate tables of corrections produced by 
smoothing results of analysis of observations, and finally compute the product 
of rotational matrices that depend on these intermediate quantities.

  At first glance it seems that the complexity of representing the Earth's 
rotation is unavoidable, since it should reflect the complexity of a theory 
for modeling the Earth's rotation at a level of the accuracy of observations, 
i.e. on the order of $10^{-10}$~rad\footnote{SI units are used throughout 
the paper. Conversion factors to non-standard units: $ 1 \cdot 10^{-9} 
\mbox{rad} \approx 0.21 \mbox{mas} \approx 14 \mu\mbox{sec}$}. This would be 
true if theoretical models were precise to that level. Roughly speaking, 
the Earth's rotation can be considered as consisting of two 
components, the tidally driven component with precisely known frequencies
and the component driven by an exchange of the angular momentum between 
the solid Earth and geophysical fluids. The latter component is not 
predictable in principle. The atmosphere contributes to the UT1 at a level 
of $10^{-6}$ rad, more than three orders of magnitude higher than the 
accuracy of observations. For a long period of time it was considered that 
the quasi-diurnal motion of the pole, namely the precession and nutation, 
can be modeled with a precision comparable to the accuracy of observations. 
First results of VLBI analysis presented by \citet{r:Her86} shattered this 
belief. It was discovered that even the most advanced nutation theory 
of \citet{Wahr80} was not accurate enough. Numerous attempts to build 
a theory of tidally driven quasi-diurnal motion were made, but they were 
not completely successful. In order to reduce the disagreement between 
theories and observations, the authors had to resort to adjusting some 
parameters of their theories to quantities derived from \emph{the same 
observations} of the quasi-diurnal motion 
\citep{r:Mat91,r:mhb2000,r:Getino2001}. It was also soon realized that 
there are two other constituents of the quasi-diurnal polar motion at a level 
of 1~nanoradian: the free core nutation \citep{r:moritz} and the nutation 
excited by the atmosphere \citep{r:Biz98}. These constituents currently 
cannot be predicted and, presumably, they cannot be predicted in principle. 
Therefore, even if a precise theory of forced nutation is built in 
the future, one should apply parameters determined from observations in 
order to represent the quasi-diurnal motion. 

  Recognizing that both components in the Earth rotation cannot be 
predicted with accuracy comparable to the precision of observations,
prompts us to reconsider approaches to representing the Earth's 
rotation and the role of the theory. The quasi-diurnal motion should be 
described with the use of parameters determined by continuous observations 
in a similar way as the UT1 and Chandler polar motion. The theory of nutation 
should be considered not as a tool for data reduction or for predicting the 
Earth's orientation, but as means for validating geophysical models. At the 
same time a theory of the Earth's rotation can provide valuable guidance 
for building empirical mathematical models.

  The goals of this paper is to build such an empirical mathematical model,
to demonstrate using a long dataset of observations that it is feasible
and to show that this approach describes the Earth rotation at least as well 
as the traditional way. It should be noted that an empirical mathematical model 
has a different meaning than a theoretical model. The theoretical model 
relates a function of time that describes the Earth rotation with specific 
properties of the Earth's body in the form of a solution for the equation of 
dynamics. The empirical mathematical model relates this function of time 
to observations using a parameter estimation technique.  A minimal requirement 
for an empirical model is to represent the phenomena with the least possible 
errors for the entire interval of observations and to provide estimates of 
uncertainties. We will also try to satisfy two additional requirements: 
the model should be simple and the parameters of the model should not be 
strongly correlated.  If parameter estimates are strongly correlated, 
their comparison with theoretical predictions becomes problematic. We will 
represent the Earth orientation parameters (EOP) in the form of an expansion 
over a family of basis functions. 

  The procedure for developing an empirical model of the Earth's rotation 
is presented in the rest of the paper. The choice of the a~priori 
model and basis functions is described in section~\ref{s:choice},
the proposed mathematical models is described in Sect.~\ref{s:basis},
the strategy of analysis of the 22 year long dataset of VLBI observations 
is presented in Sect.~\ref{s:anal}, and the results of solutions are discussed 
in Sect.~\ref{s:res}. Concluding remarks are given in Sect.~\ref{s:conclude}.

\section{Choice of the a~priori model and the basis functions}
\label{s:choice}

\subsection{Model of observations}

  We consider here that $N$ stations observe $K$ celestial physical bodies. 
It is assumed that each station is associated with a reference point. 
In the case of VLBI antennas with intersecting axes, this is the intersection 
point of the axes. Observing stations receive electromagnetic radiation 
emitted by celestial bodies, and each sample of the received signal 
is associated with a time stamp from a local frequency standard synchronized 
with the GPS time. Analysis of voltage and time stamps of received radiation 
eventually allows us to derive photon propagation time from reference 
points of observed bodies to reference points of observing stations. 
These distances depend on the relative positions of stations with respect 
to the observed bodies. The instantaneous coordinate vector of station 
$i$, $\vec{r}_i(t)$, at a given moment of time is represented as the
sum of a rotation and translation applied to a vector $\vec{r}_i(t_0)$ 
at initial epoch $t_0$ as
\begin{eqnarray}
   \vec{r}_i(t) = \widehat{\mathstrut\cal M}(t)\, \vec{r}_i(t_0) + \vec{T}(t) + 
                  \vec{d}_i(t)
\label{e:e1}\end{eqnarray}
  where $ \widehat{\mathstrut\cal M} $ is the rotation matrix, $\vec{T}(t)$ is
the translational motion of the network of stations, and $\vec{d}(t)$ 
is a displacement vector. Equations of photon propagation tie the 
instantaneous vector of site coordinates $\vec{r}_i(t)$ with vectors 
of observed physical bodies and their time derivatives. These relationships 
allow us to build a system of equations of conditions.  Station position 
vectors at a given epoch and the quantities on the right-hand side of 
expression \ref{e:e1} are estimated from a single least square solution.

\begin{figure}
   \resizebox{\hsize}{!}{\includegraphics[angle=-90]{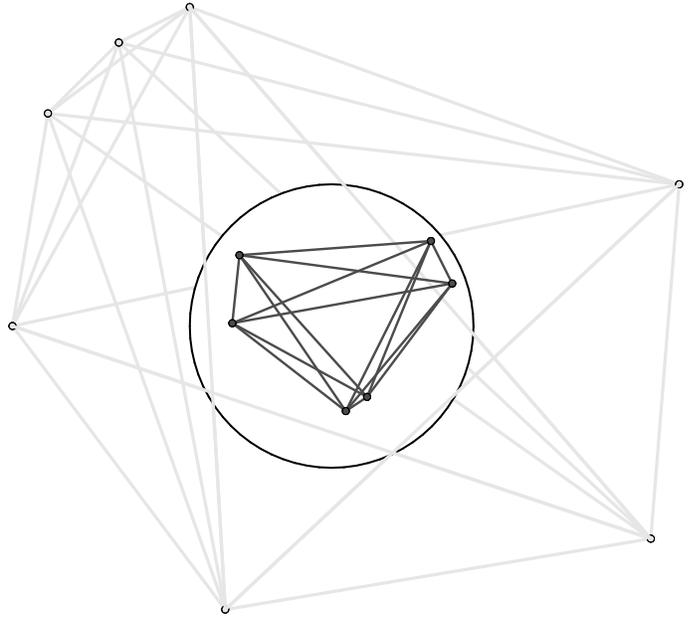}}
   \caption{The polyhedron of observing stations (black) and the polyhedron 
            of observed bodies (grey). The relative orientation of two 
            polyhedrons is estimated from observations of projections of
            vectors between observing stations and observed bodies and 
            interpreted as the Earth's rotation. }
   \label{f:polyhedron}
\end{figure}

  The displacement vector $\vec{d}_i(t)$ characterizes the motion of an
individual station, while matrix $ \widehat{\mathstrut\cal M} $ and
vector $\vec{T}$ describe the motion of the entire network. Assuming the 
stations are solidly connected to the Earth's crust, we consider that this 
part of motion represents the motion of the entire Earth. In particular, 
matrix $ \widehat{\mathstrut\cal M}(t) $ describes the Earth's rotation. 
Schematically, the mechanical model of observations can be viewed as the
motion of the polyhedron of observing stations with respect to the polyhedron
of observed bodies (Fig.~\ref{f:polyhedron}). It should be noted that the
EOP are defined here as the parameters of an estimation model, while in the 
framework of the traditional approach, they are defined as angles between
big circles on a sphere. 

  Since both $\widehat{\mathstrut\cal M}(t)$ and vector $\vec{d}(t)$ are 
functions of time, i.e. infinite sets of points, they can be evaluated from
a finite set of observations only in the form of an expansion in some functions. 
When we say that the matrix $\widehat{\mathstrut\cal M}(t)$ is determined 
from observations, this should not be understood literally, but instead it 
should be construed that a mathematical model for the dependence of 
$\widehat{\mathstrut\cal M}(t)$ on time is assumed, either explicitly or 
implicitly. The model depends on a finite set of unknown parameters that
are determined from observations. 

  The choice of the mathematical model is not unique. On one hand, the 
mathematical model should approximate the rotation with errors comparable 
to uncertainties of observations during the full interval of observations. 
On the other hand, we should be able to estimate robustly all the parameters 
of the model. Let us consider several approaches.

\subsection{The time series approach}

  The easiest way to represent a rotation is to estimate the matrix 
$ \widehat{\mathstrut\cal M}(t) $ at certains moment of time and, thus, 
generate the time series. The $ 3\times 3 $ matrix 
$ \widehat{\mathstrut\cal M} $ has 9 elements, but only three of them are 
linearly independent. An arbitrary rotation matrix can be decomposed in
a product of several elementary rotation matrices with respect to coordinate 
axes at certain angles. Therefore, it is sufficient to determine these rotation 
angles in order to determine the matrix $ \widehat{\mathstrut\cal M}(t) $ 
from observations.

  The fundamental problem is that no observation technique, except the laser
gyroscope, is sensitive to the instantaneous Earth's rotation vector or to
its time derivatives \emph{directly}. The rotation angles can be derived 
using the least square estimation procedure, together with evaluating other 
parameters. It requires accumulating sufficient amount of data in order 
to separate variables. The estimates of the Earth's rotation angles cannot 
be sampled too fast. A typical sampling rate of estimates is one day, 
since this allows compensation for a certain type of systematic error. 
In some cases the sampling rate can be reduced to several hours.

  Unfortunately, one cannot neglect changes in the Earth's rotation angles 
during the sampling interval.  The accuracy in determining rotation angles 
for the 24-hour period is nowadays at the level of $ 2\mbox{--}5 \cdot 
10^{-10} $ rad. The amplitude of the quasi-diurnal motion around axes~1 and 2 
is growing with a rate that is an order of magnitude of 
$7 \cdot 10^{-12}$ rad s${}^{-1}$. Therefore, this motion should first be 
separated from the slowly varying components. In the era of optical 
astrometry, some components of this motion, namely precession and nutation, 
were determined separately from observations of slowly varying components 
using a different technique and even different instruments. The observations 
of slowly varying constituents in the Earth's rotation angles were corrected 
with a model of the quasi-diurnal motion. \citet{r:Her86} have demonstrated 
that corrections to the model of the quasi-diurnal motion around axes 1 and 2 
can be estimated together with slowly varying components of the Earth rotation, 
if the rotation angles around coordinate axes $A_i(t)$ are parameterized 
in the form 
\begin{eqnarray}
   \begin{array}{l @{\:} l @{\:} l @{\;} l @{\;} l @{\;} 
                 l @{\;} l @{\;} l @{\;} l @{\;} l @{\;} l}
      A_1(t) & = & b_1(t) & + & \dot{b}_1(t)(t-t_0) 
                 & + & c(t) \, \cos \! - \Omega_n t 
                 & + & s(t) \, \sin \! - \Omega_n t                   \\
      A_2(t) & = & b_2(t) & + &  \dot{b}_2(t)(t-t_0) 
                 & + & c(t) \, \sin \! - \Omega_n t  
                 & - & s(t) \, \cos \! - \Omega_n t                   \\
      A_3(t) & = & b_3(t) & + & \dot{b}_3(t)(t-t_0) &  & & &  &  &    \\
   \end{array}
\label{e:e2}\end{eqnarray}
  where $\Omega_n$ is the nominal diurnal Earth's rotation rate, 
\mbox{$ 7.292\,115\,146\,706\,707 \cdot 10^{-5} \: \mbox{rad} \; \mbox{s}^{-1}$}.
Parameters $c(t)$, $s(t)$, $b_i(t)$ are slowly varying functions of time. 
This approach quickly became traditional for processing VLBI experiments, 
and eight parameters, $b_1, b_2, b_3, \dot{b}_1, \dot{b}_2, \dot{b}_3, c, s $ 
are routinely determined for each individual 24~hour observing session.

\subsection{Limitations of the time series approach}

  However, it is important to realize the limitations of the time series 
approach. First, the raw time series of estimates provides the values of 
rotation angles only at specific discrete moments. They do not determine 
a functional dependence of rotation angles on time. An end user needs to have 
a tool for computing Earth's orientation at any moment of time within 
the interval of observations. Thus, the raw time series are the basis for the 
second step of processing that involves smoothing and interpolation. Smoothing 
and interpolation of the time series $c_k, s_k, b_{1k}, b_{2k}, b_{3k} $ 
implicitly assumes that $A_i(t)$ satisfies some mathematical model that
appears to be different from the model in expression~\ref{e:e2} used in the 
estimation process. The resulting smooth function of rotation angles does 
not provide the best fit to observations; if it did, no smoothing would have 
been needed.

  Second, at the present level for accuracy of observations, the estimation 
model corresponding to Eqs.~\ref{e:e2} is not adequate: one cannot neglect 
changes in $c(t), s(t)$ and $\dot{b}_i(t)$ over the interval 
of estimation, typically 24 hours.  Adjusting time derivatives $\dot{c}(t), 
\dot{s}(t), \ddot{b}_i(t)$ makes estimates of these parameters so strongly 
correlated that they do not have a practical value. 

   Changes in the a~priori model for slowly varying components of rotation 
angles affect all estimated parameters, including $c(t)$ and $s(t)$. In order 
to demonstrate this, two VLBI solutions using 3563 twenty four hour observing 
sessions from 1984 through 2006 were computed. The USNO Finals EOP time series 
of pole coordinates and UT1--TAI with a time span of 
1~day~\citep{r:iers-anrep}\footnote{Available on the Internet at 
ftp://maia.usno.navy.mil/ser7/finals.all} were used as the a~priori model
in the reference solution. The Gaussian noise with standard deviation 
1~nrad was added to all components of the USNO Finals EOP series, and these 
modified time series were used in the trial solution. The rms of differences 
in the \emph{total values} of $b(t), c(t), s(t)$, i.e., the sum of a~priori 
values and adjustments over the 24 hour time intervals, was 
0.14~nrad for $b(t)$ and 0.16~nrad for $c(t) and s(t)$. Since the accuracy 
of estimates of $b(t), c(t), s(t)$ from 24~hour VLBI experiments is currently 
at a 0.3~nrad level, the accuracy of the a~priori EOP series should be better 
than 1~nrad in order to reduce the contribution its errors to a negligible 
level: 1/2 of the random error in estimates. In a similar way, the change 
in the a~priori model for the quasi-diurnal motion also affects estimates of 
$c(t), s(t)$ and $b_i(t)$. Although one can expect that 
a continuous process or refining the a~priori model and subsequent least 
square estimation should converge, this does not happen in practice. It is 
known among analysts who process raw data that, if the initial a~priori values 
are changed, the total angles, i.e. the sum of the a~priori and the 
adjustments, come out different. Researchers who process time series are not 
always aware of these complications and tend to consider the results 
of processing the same observations by different analysts as independent 
``data'', so they attribute the differences between them to so-called 
``analyst noise''. These discrepancies occur due to an internal inconsistency 
between the estimation model, the a~priori model, and the post-processing 
procedure. 

   Third, the second step in the analysis, smoothing and interpolation, 
is rather subjective. A different degree of smoothing produces a different 
series. 

   Finally, the time series cannot be used directly for making an inference 
about the physical processes that affect the Earth's rotation. The time series
are transformed by various analysis procedures. The dependence of the 
series on the a~priori model and the correlations between the elements of 
time series are usually ignored. The correlations between the elements are 
not zero, because the elements themselves were estimated together with other 
parameters like global site velocity or source positions that are considered
common for the entire interval of observations. Although these correlations are 
not strong, typically at a level of 0.1, their contribution is significant when
long time series are processed.

  Due to the complexity of the a~priori model, analysts who process the time 
series of estimates usually do not handle the total angles of the Earth's 
rotation, but rather adjustments to the a~priori values, tacitly assuming that 
analysts who processed the raw data strictly followed a standardized procedure for data 
reduction. In reality this is often not the case. This creates an additional 
source of confusion and errors. 

  These complications prompt us to look for a one-step procedure of 
estimation of the Earth's orientation parameters.

\section{Representation of the Earth rotation in the form of the expansion 
         into basis functions}
\label{s:basis}

  While Eqs.~\ref{e:e2} represent rotation angles within a short period
of time, 24 hours, they are not adequate for a longer period of time. We need
to find a mathematical model which would be valid for the entire period 
of observations, i.e. several decades. The matrix 
$\widehat{\mathstrut\cal M}_a$ has a non-linear dependence on its 
arguments. A linear estimator, the least square method, allows us to evaluate 
not the matrix itself, but its small perturbation. The coordinate 
transformation of a vector $\vec{r}$ from the terrestrial coordinate system 
to the celestial coordinate system is then written as 
\begin{eqnarray}
   \vec{r}_{{}_C} = \widehat{\mathstrut\cal M}_a(t) \, \vec{r}_{{}_T} +
                    \bigl( \vec{q}_e(t) + \vec{q}_a(t) \bigr) \times 
                    \vec{r}_{{}_T}
\label{e:3}\end{eqnarray}
where $\vec{r}_{{}_C}$  and $\vec{r}_{{}_T}$  designate the coordinates of 
the vector $\vec{r}$ in the celestial and terrestrial coordinate systems, 
respectively, $\vec{q}_e(t)$ is the vector of a small perturbational rotation, 
$\vec{q}_a(t)$ is the small a~priori rotation vector in the terrestrial 
coordinate system, and $ \widehat{\mathstrut\cal M}_a(t) $ is the a~priori 
matrix of finite rotation. Vectors $\vec{q}_e(t)$ and $\vec{q}_a(t)$ 
are small in the sense that we can neglect squares of their components. 
The vector $\vec{q}_a(t)$ can be set to zero by an appropriate choice of 
the matrix $\widehat{\mathstrut\cal M}_a(t)$. Considering that the accuracy 
of determination of rotation angles averaged over a 24 hour period is currently
at the level of 3$\,\cdot 10^{-10} $ rad, and the accuracy of estimates of 
amplitudes of harmonic constituents averaged over the period of 20 years is at 
the level of $ 10^{-11} $ rad, the components of vectors $\vec{q}_a(t), 
\vec{q}_e(t)$ should not exceed $ 3 \cdot 10^{-6} $ rad.  It should be noted 
that these requirements on accuracy of the a~priori model are three orders
of magnitude weaker than those needed for an unbiased estimation of time series. 

\begin{figure}
   \raisebox{0.06\textwidth}{\includegraphics[width=0.13\textwidth,angle=90]{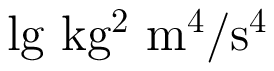}}
   \includegraphics[width=0.45\textwidth]{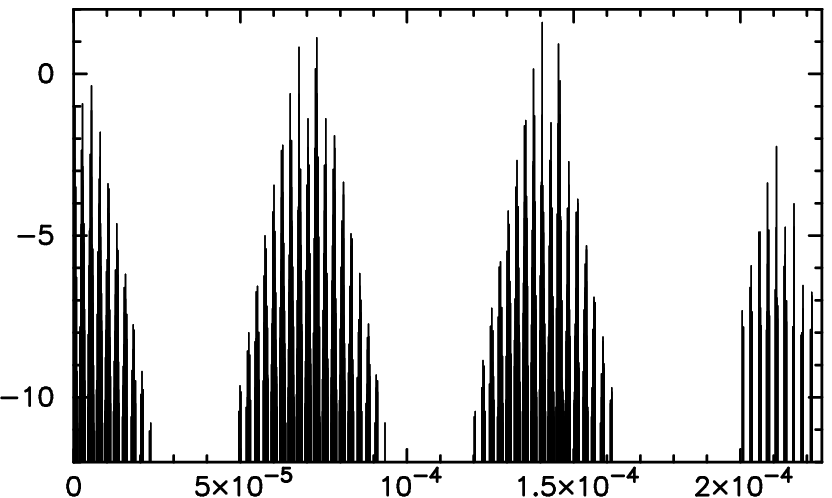}
   \par\vspace{-1ex}
   \parbox{0.49\textwidth}{\hfill rad s${}^{-1}$}
   \caption{The logarithm of the power spectrum of the tide generating 
            potential in $\mbox{kg}^2 \mbox{m}^4 \mbox{s}^{-4}$ as 
            a function of the angular frequency.}
   \label{f:tidspe}
\end{figure}

  In order to find an appropriate basis for expanding of $\vec{q}_e(t)$, we
need to use an a~priori knowledge of the process under consideration.
The Earth's rotation can be considered in terms of a response to external 
forces. The external forces that affect rotation of the solid Earth are 
caused 1) by redistribution of geophysical fluids and 2) by tidal attraction 
of external bodies. The first process is not predictable and 
is dominating at frequencies by modulo much less than the diurnal frequency
$\Omega_n$. The tide-generating potential exerted by external bodies can 
be considered to be known precisely. Its spectrum has a comb of very sharp 
lines as shown in Fig.~\ref{f:tidspe}.

  To characterize the Earth's response, we should take into account 
that the triaxiality of the Earth's inertia tensor $(B-A)/C$ is small, 
about $2 \cdot 10^{-5}$. Therefore, the differential equations of the Earth's 
rotation are linear. First, this leads to decoupling rotation around the axes 
1 and 2, i.e. the polar motion, and rotation around the axis 3, the diurnal 
motion. Second, the response to harmonic external forces will result 
in harmonic variations of the component 2 of $\vec{q}_e$ with the same 
amplitude as component 1 with the phase shifted by $-\pi/2$. Third,
the excitation at the diurnal frequency will result in the appearance of 
cross-terms $t \sin \! - \Omega_n$ and $t \cos \! - \Omega_n$ \citep{r:moritz}. 

  Considering these factors, the following mathematical model for the 
the vector of a small perturbational rotation is proposed:
\begin{eqnarray}
   \vec{q}_e(t) = \left(
     \begin{array}{ l@{\;} l@{\;} l}
       \displaystyle\sum_{k=1-m}^{n-1} f_{1k} \, B_k^m(t) \:&  + &
       \displaystyle\sum_{j}^{N} \left( P^c_{j} \cos \omega_m \, t \: + \:
                                        P^s_{j} \sin \omega_j \, t \right)   
       \vspace{0.5ex} \\
       & + & t \, \left( S^c \cos \! - \Omega_n \, t \: + \:
                         S^s \sin \! - \Omega_n \, t \right) 
       \vspace{0.5ex} \vspace{2ex} \\
       \displaystyle\sum_{k=1-m}^{n-1} f_{2k} \, B_k^m(t) \: & + & 
       \displaystyle\sum_{j=1}^{N} \left( P^c_{j} \sin \omega_j \, t \: - \:
                             P^s_{j} \cos \omega_j \, t \right)
       \vspace{0.5ex} \\
       & + & t \, \left( S^c \sin \! - \Omega_n \, t \: - \:
                         S^s \cos \! - \Omega_n \, t \right)
       \vspace{0.5ex} \vspace{2ex} \\
       \displaystyle\sum_{k=1-m}^{n-1} f_{3k} \, B_k^m(t) \: & + &
       \displaystyle\sum_{j=1}^{N} \left( E^c_{j} \cos \omega_j \, t + 
                             E^s_{j} \sin \omega_j \, t \right) 
       \\
     \end{array}
     \right)
\label{e:4}\end{eqnarray}
  where $B_k^m(t)$ is the B-spline function of degree $m$ determined at a 
sequence of knots $ t_{1-m}, \, t_{2-m}, \, \ldots, \, t_0, \, t_1, \, 
\ldots \, t_k$; $\omega_j$ are the frequencies of external forces;
the coefficients $f_{ik}, P^c_j, P^s_j, S^c, S^s, E^c_{j}, E^s_{j}$; 
are the parameters of the expansion, $\Omega_n$ is the nominal frequency 
of the Earth's rotation. Here $n$ is the dimension of the B-spline basis 
and $N$ is the dimension of the Fourier basis. Thus, the vector of small 
perturbational rotation is expanded into the basis of B-splines, which 
is orthogonal over the entire period of observations, and the basis of 
harmonic functions, which is orthogonal in the range $(-\infty, +\infty)$. 
The first basis approximates the slowly varying component in the Earth's 
rotation, the second basis approximates the quasi-diurnal component, as 
well as other harmonic constituents of the Earth's rotation. 

\subsection{The B-spline basis}

  The B-spline basis functions were first introduced by \citet{r:Schoe}. The
B-spline function of degree $m$ depends on time and on a monotonically 
nondecreasing sequence of $n$ knots at the interval $[t_1, t_n]$. In order
to introduce splines, let us extend this sequence by adding $m$ elements at
the beginning of the sequence and $m$--1 elements at the end of the sequence
such that $ t_{1-m} \! = \! t_{2-m} \! = \, \ldots \, \! = \! t_0 \! = \! t_1 $
and $ t_n \! = \! t_{n+1} \! = \! t_{n+2} \, = \, \ldots \, = \, t_{n+m-1} $. 
At a given 
extended sequence of $n+2m-1$ knots, $n+m-1$ B-spline functions with 
the pivot element $k \in 1-m, \, 2-m, \, \ldots \, n-1$ are defined through 
a recursive relationship.

  The B-spline of the 0-th degree with the pivot knot $k \in [1,\, n \! - \! 1]$ 
on the knots sequence \mbox{$ (t_1, \, t_2, \, \ldots \, , t_{n} )$}, such that 
$ t_1 \leq t_2 \leq \, \ldots \, \leq t_{k}$, is determined by
\begin{eqnarray}
     B_k^{0}(t) = \left\{ \begin{array}{ll} 
                                 1,  & \mbox{if} \quad t \in [t_k, t_{k+1})  \\
                                 0,  & \mbox{otherwise}
                           \end{array} 
                  \right. .
\label{e:5}\end{eqnarray} 
  The B-spline of the $m$th degree with the pivot knot 
$k \in [ 1 \! - \! m, \, n \! - \! 1]$ on the extended sequence of knots 
$ (t_{1-m}, \, t_{2-m}, \, \ldots \, t_{n+m-2}, \,  t_{n+m-1})$ is expressed 
via the B-splines of the $ m \! - \! 1$th degree as
\begin{eqnarray}
     B_k^m(t)= \Frac{t - t_k}{t_{k+m} - t_k} \, B_k^{m-1}(t) \: + \:
               \Frac{t - t_{k+m+1}}{t_{k+1} - t_{k+m+1}} \, B_{k+1}^{m-1}(t)
\label{e:6}\end{eqnarray}
  
  Computation of B-splines is as simple as computation of other polynomials.
Similar simple recursive relationships exist for derivatives of B-splines
and integrals. The B-spline of degree $m$ with the pivot element $k$ is 
non-zero only at the interval $(t_{k}, t_{k+m+1})$. It can be proved that a
sequence of $n+m-1\:$ B-spline functions of degree $m$ with pivot elements
$k \in 1-m, \, 2-m, \ldots, n-1\:$ forms a basis on the interval $[t_1, t_n]$.
The proof of this and many other useful theorems related to B-splines
can be found in \citet{r:Nue}.

  In general, knots can be selected arbitrarily. Test runs have shown that a set
of B-spline functions of the 3rd degree with equidistant knots with a time 
span of 3 days for components~1, 2, and 1 day for component~3 of the vector 
$\vec{q}_e(t)$ adequately represents the slowly varying component of 
the Earth's rotation. Weak constraints on values of B-splines, its first
and second derivatives can be imposed to ensure the stability of the solution
at intervals with considerable gaps in observations and at the beginning
and the end of the data set.

\subsection{The Fourier basis}

\begin{figure}
   \raisebox{0.06\textwidth}{\includegraphics[width=0.13\textwidth,angle=90]{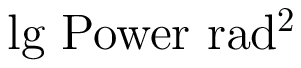}}
   \includegraphics[width=0.45\textwidth,clip]{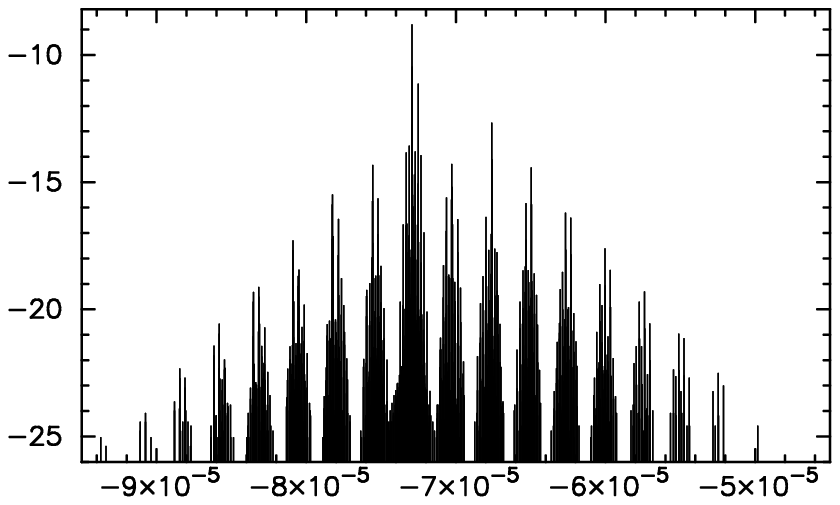}
   \par\vspace{-1ex}
   \parbox{0.49\textwidth}{\hfill rad s${}^{-1}$}
   \caption{The logarithm of the power spectrum of quasi-diurnal variations 
            in $q_1, q_2$ components of the rotation vector according to 
            the REN--2000 expansion in $\mbox{rad}^2 $ as a function of the 
            angular frequency.}
   \label{f:nutspe}
\end{figure}

  Modeling the quasi-diurnal components is more challenging. The tides exerted
by the Moon and the Sun cause variations in sea currents and the sea 
surface at tidal frequencies. These variations excite changes in all 
components of the Earth's rotation. The resonance near the retrograde diurnal 
frequency causes a significant amplification at that frequency band for 
components 1 and 2 of the vector $\vec{q}_e(t)$. The theoretical spectrum 
of this motion referred to as nutation computed by 
\citet{r:ren2000a,r:ren2000b} and \citet{r:ren2000c} for the model of the 
rigid Earth, the REN--2000 expansion, is presented in Fig.~\ref{f:nutspe}.

  The problem is that the spectrum is very dense, and observations during
a finite period of time cannot resolve all the constituents. The REN--2000 
spectrum has 560 constituents with amplitudes greater than $10^{-11}$ rad 
and 1551 constituents with amplitudes greater than $10^{-12}$ rad with 
the frequency difference between some of them as low as $10^{-15}$ 
\mbox{rad s${}^{-1}$}. 

  Several strategies can be used for overcoming this problem. First,
we can select frequencies with maximal amplitudes from the theoretical
spectrum and ignore constituents with an angular frequency separation 
less than $\omega_{min} = 2 \pi/\Delta T$, where $\Delta T$ is the interval
of observations. No signal will be mismodeled if the frequency separation 
between the constituents $ \Delta \omega \ll \omega_{min}$, since in this 
case the two constituents will be indistinguishable. 

  However, if the constituents are not very close, the mismodeled signal will 
leak into adjustments at other frequencies. The sidelobe with the amplitude 
$A_2$ and frequency $\omega_2$ of the main constituent with the 
frequency $\omega_1$ can be omitted if the quantity
$ A_2\frac{\displaystyle\strut \omega_2 - \omega_1}
          {\displaystyle\strut \Delta \omega_{min}}$ is less than a certain 
threshold. Depending on the threshold level, there are several hundred
constituents in the tidal spectrum for which this condition is not valid.

   One way to mitigate this problem is to estimate the amplitude of close 
sidelobes, together with the amplitude of main constituents, and to impose
strong constraints on the amplitude of sidelobes by using some a~priori
information. It is plausible to assume that the a~posteriori amplitudes
of constituents of the quasi-diurnal motion for the real Earth differ from 
the theoretical amplitudes computed for the rigid Earth by multiplicative 
factors called transfer function, which is a smooth function of frequency
according to theory.  Then we can assume that the transfer function for two 
close constituents with theoretical amplitudes $P$ be the same, i.e. 
the ratio of complex amplitudes $A$ of two close constituents 
is the same as for the a~priori rigid Earth amplitudes, and, therefore, 
should satisfy this equation:
\begin{eqnarray}
\frac{\displaystyle\strut P^c_1 + i \, P^s_1}
     {\displaystyle\strut P^c_2 + i \, P^s_2} = 
\frac{\displaystyle\strut A^c_1 + i \, A^s_1}
     {\displaystyle\strut A^c_2 + i \, A^s_2} \quad .
\label{e:e7}\end{eqnarray}

  Although this approach reduces the leakage from a mismodeled signal, it is not
fully satisfactory. In general, using strong constraints is undesirable, since
this introduces a bias in estimates. The validity of Eq.~\ref{e:e7} 
cannot be confirmed or refuted from observations. It comes from a theory. But 
if the estimation model implicitly incorporates theoretical assumptions, 
strictly speaking the estimates cannot be used for validation of the theory. 
Although Eq.~\ref{e:e7} for EOP variations caused by the tidal potential 
exerted by external bodies has a sound theoretical basis, we should bear 
in mind that the ultimate goal of comparing  theoretical predictions with 
observations is to check the validity of assumptions built into the foundation 
of the theory and to make a judgment whether the model is complete or not. 
If there are unaccounted additive constituents at these frequencies, 
for example, caused by the free motion, by the atmospheric, or by oceanic 
excitation, the Eq.~\ref{e:e7} may not be valid.

  An alternative to constraining sidelobes is the strategy of estimating 
a wide  range of constituents that are multiples of $\omega_{min}$ or, in
other words, indirectly performing the discrete Fourier transform of
the perturbational rotation. With this approach, in general we are 
in a position to discard our a~priori knowledge about the frequency structure 
of the signal. Estimating the signal at discrete frequencies that are 
multiples of $\omega_{min}$, from the zero frequency through the Nyquist 
frequency, recovers any signal according to the sampling theorem. However, 
this kind of approach applied to estimating the vector $\vec{q}_e(t)$ 
has a practical value only if the number of non-negligible constituents in 
the discrete spectrum is significantly less than the total number of samples. 
Since the spectrum of the tide-generating potential consists of a set of 
discrete frequencies that are not commensurate to each other, the frequencies
that are multiples of $\omega_{min}$ cannot coincide with all tidal 
frequencies.  If the amplitude of a narrow-band harmonic signal is not 
estimated at its frequency, but estimated at a set of nearby frequencies 
that are multiples of $\omega_{min}$, the signal will be recovered only 
partially. The wider the range of frequencies, the better the approximation. 
The rate of convergence depends on the amplitude of the signal and the 
difference between its frequency and the closest frequency used for estimation. 
Selection of reasonably good a~priori $\vec{q}_a(t)$ values may significantly 
reduce the number of frequencies needed for estimation to reach a given 
level of accuracy of approximation.

  Other important constituents of the signal at the retrograde diurnal
band are the free near-diurnal wobble \citep{r:moritz} and the atmospheric 
nutations \citep{r:Biz98,r:Yse02}. Since this signal is excited by 
a broad-band stochastic process, it is expected that these constituents in the 
Earth's rotation are also relatively broad-band. To model this 
signal, the constituents at frequencies within the range of that band need
to be added to the list of constituents at tidal frequencies. It follows 
from the sampling theorem of \citet{r:Kot33} that a band limited signal with 
frequencies in the range of $[\omega_l, \omega_h]$ is completely recovered 
when the estimates of the sine and cosine amplitudes of the spectrum are 
made at discrete frequencies
$[\omega_l,\: \mbox{$\omega_l + \omega_{min}$}, \enskip 
              \mbox{$\omega_l + 2\: \omega_{min}$}, \enskip  \ldots \enskip 
              \mbox{$\omega_l + (N-1)\, \omega_{min}$}, \enskip  \omega_h]$ .

   The tidal spectrum also has constituents with low frequencies, so-called
zonal tides. They affect component~3 in the vector of the perturbational
rotation. Their contribution dominates the rate of change for this component.
It would be desirable to estimate the complex amplitude of this variations. 
Since the residual rate of change of $\vec{q}_e$ is a factor of 3--10 less, 
constraints on a rate of change for the residual component of $\vec{q}_e$,
modeled with an expansion over the B-spline basis can be set stronger without
introducing a bias in the estimates. This improves the solution stability during 
intervals of time with gaps in observations. For the same reason, it would 
be desirable to estimate variations in components~1 and 2 of the Earth's 
rotation vector at the annual and Chandler frequencies: 
$1.990968 \cdot 10^{-7}$ and $ 1.678 \cdot 10^{-7} $ rad s${}^{-1}$ ,
respectively.

\subsection{Decorrelation constraints}

  It should be noted that the estimates of harmonic constituents with lower 
periods than the time span between nodes of B-spline will so highly correlate 
with B-spline coefficients that the system of equations will be very close 
to singular. Decorrelation constraints on coefficients of the B-spline should 
be imposed in order to overcome this problem. We require that the product 
of expansion over basic B-spline and Fourier functions for the $j$th 
frequency be zero over the interval of observations:
\begin{eqnarray}
   \begin{array}{lcr}
       \displaystyle\int\limits_{t_0}^{t_1} \left(
          \sum_{k=1-m}^{n-1} f_k \, B_k^m(t) \: \cdot \:
          \displaystyle\sum_{j=1}^{N} P^c_{j} \cos \omega_j \, t \right) \: dt  
           = 0\\
       \displaystyle\int\limits_{t_0}^{t_1} \left(
          \sum_{k=1-m}^{n-1} f_k \, B_k^m(t) \: \cdot \:
          \displaystyle\sum_{j=1}^{N} P^c_{j} \sin \omega_j \, t \right) \: dt  
          = 0\\
   \end{array} \quad .
\label{e:e72}\end{eqnarray}
  
  This is reduced to 
\begin{eqnarray}
   \begin{array}{lcr}
     \displaystyle\sum_{k=1-m}^{n-1} f_k \int\limits_{-\infty}^{+\infty} 
                   B_k^m(t) \cos \omega_j \, t \: dt  = 0 \\
     \displaystyle\sum_{k=1-m}^{n-1} f_k \int\limits_{-\infty}^{+\infty} 
                   B_k^m(t) \sin \omega_j \, t \: dt  = 0 
   \end{array} \quad .
\label{e:e74}\end{eqnarray}

   Thus, two constraint equations for each frequency are to be imposed.
The Fourier integral of a B-spline of the $m$th degree in Eq.~\ref{e:e74}
on a knots sequence \mbox{$ (t_k, t_{k+1}, \, \ldots \, , t_{n} )$} with 
the pivot knot $k$ such that $ k-m \leq n-1 $ is expressed through the Fourier
integral of a B-spline of the $m-1$ th degree:
\begin{eqnarray}
  \begin{array}{ll}
    F_k^m(\omega) = \\
    \displaystyle \int\limits_{-\infty}^{+\infty} B_k^m(t) \, 
                  e^{i \, \omega \, t} \, dt = 
    -\Frac{i}{\omega} \biggl( B_k^m(t_n) \, e^{i \, \omega \, t_n} -
                             B_k^m(t_1) \, e^{i \, \omega \, t_1} \biggr) 
    \:\: + \: \\ 
    \Frac{i \, m}{\omega (t_{k+m} - t_k)}        F_k^{m-1}(\omega) +
    \Frac{i \, m}{\omega (t_{k+1} - t_{k+m+1})}  F_{k+1}^{m-1}(\omega) 
    \end{array} \quad .
\label{e:e76}\end{eqnarray}

  The Fourier integral of a B-spline of the 0-th degree 
on the same sequence \mbox{$ (t_k, t_{k+1}, \, \ldots \, , t_{n} )$} 
with the pivot knot $i$ is
\begin{eqnarray}
  F_k^0(\omega) = \displaystyle \int\limits_{-\infty}^{+\infty} B_k^0(t) \, 
                 e^{i \, \omega \, t} \, dt = 
                 \Frac{i}{\omega} \biggl( e^{i \, \omega \, t_k} - 
                                          e^{i \, \omega \, t_{k+1}} \biggr) 
  \quad .
\label{e:e78}\end{eqnarray}

\section{Analysis of VLBI observations}
\label{s:anal}

\subsection{The VLBI dataset}

  A set of estimates of group delays at frequency bands centered
around 2.2 and 8.6~GHz from January 1984 through January 2006 was used to
validate the proposed approaches. The International VLBI Service for 
Geodesy and Astrometry (IVS) \citep{r:ivs} provides online access to the 
collection of all observations made in the geodetic mode under various 
astrometric and geodynamics programs from 1979 through now at 
\mbox{\tt http://ivscc.gsfc.nasa.gov}. The VLBI data set shows a substantial 
spatial and time inhomogeneity. Typically, observations are made in sessions 
with a duration of about 24 hours. Observations were sporadic in the 
early 80s, but in January 1984 a regular VLBI campaign for the determination 
of the Earth orientation parameters started first with 5-day intervals, 
from May 1993 with weekly intervals, and from 1997 twice a week. 
In addition to these observations, various other observing campaigns were 
running. On average, 150 sessions per year have been observed since 1984.

  During that period 153 stations participated in observations, although 
a majority of them observed only during short campaigns. The observations 
at stations that participated in less than $20\,000$ observations, and the
stations that only participated in at regional networks with sizes of 2000~km 
and less were discarded. Forty four stations remained. Observations of 
sources that were observed in less than 4 sessions and gave less than 
64 usable pairs of dual-band group delays were excluded. The data before 
January 1984 were also discarded. In total, $ \sim $5\% of the observations 
were excluded, and the remaining data from 3563 sessions between January 
1984 to August 2006, more than 4.6 million of dual-band pairs of group
delays, were used in the analysis. 

  The number of participating stations in each individual session varies from 
2 to 20, although 4--7 is a typical number. No station participated in all
sessions, but every station participated in sessions with many different 
networks. All networks have common nodes and, are therefore, tied together. 
Networks vary significantly, but more than 70\% of them have a size exceeding
the Earth's radius.

\subsection{Theoretical model}

   The state of the art theoretical models were used for computing the
theoretical time delay and its partial derivatives. The procedure in general
follows the approach presented by \citet{r:masterfit} with some minor 
refinements. The expression for time delay derived by \citet{r:Kop99} in the 
framework of general relativity was used. The displacement caused by the 
Earth's tides were computed using a rigorous algorithm \citet{r:harpos} 
with a truncation at a level of 0.05~mm using the numerical values of the 
generalized Love numbers presented by \citet{r:mat01}. 
The displacements caused by ocean loading were computed by convolving the 
Greens' functions with ocean tide models using the NLOADF algorithm 
of \citet{r:spotl2}. The GOT00 model \citep{r:got99} of diurnal and 
semi-diurnal ocean tides, the NAO99 model \citep{r:nao99} of ocean zonal 
tides, the equilibrium model of the pole tide and the tide with period 
of 18.6 years were used. The atmospheric pressure loading was computed by 
convolving the Greens' functions with the numerical model of the 
atmosphere NCEP Reanalysis \citep{r:ncep}. The algorithm of computations is 
described in full details in \citet{r:aplo}.

  The a~priori path delay in the atmosphere caused by the hydrostatic 
component was calculated as a product of the zenith path delay computed
on the basis of surface pressure using the \citet{r:Saa72} expression and
the isobaric mapping function \citep{r:imf} computed using the geopotential
height of the 20~kPa pressure layer provided by the numerical weather 
model NCEP Reanalysis. The isobaric mapping function describes the dependence 
of path delay on the angle between the local axis of symmetry of the atmosphere 
and the direction to the observed sources. The direction of this axis from 
the zenith was considered to coincide with the normal to the surface of 
the geopotential height at the 20~kPa pressure level. This normal was 
computed using the NCEP Reanalysis dataset.

  Since the accuracy requirements to the of the a~priori Earth rotation model 
are very low in the framework of the present approach, we can exploit 
this to use the simplest possible model. The following expression for 
the a~priori matrix of the Earth's rotation 
$ \widehat{\mathstrut\cal M}_a(t) $ according to the Newcomb-Andoyer formalism 
was used: 
\begin{eqnarray}
   \begin{array}{ll}
      \widehat{\mathstrut\cal M}_a(t) = & 
      \widehat{\mathstrut\cal R}_3(\zeta_0)  \cdot
      \widehat{\mathstrut\cal R}_2(-\theta_0)  \cdot
      \widehat{\mathstrut\cal R}_3(z)  \cdot
      \widehat{\mathstrut\cal R}_1(-\epsilon_0)  \cdot
      \widehat{\mathstrut\cal R}_3(\Delta\psi)  \, \cdot \\ &
      \widehat{\mathstrut\cal R}_1(\epsilon_0 + \Delta\epsilon) \cdot 
      \widehat{\mathstrut\cal R}_3(-S)
   \end{array}
\label{e:e8}\end{eqnarray}
  where $ \widehat{\mathstrut\cal R}_i $ is a rotation matrix around
the axis $i$. For the variables $\zeta_0, \theta_0, z, \epsilon_0, \Delta\psi,
\Delta\epsilon_0, S, $ the following simplified expressions were used:
{\small
\begin{eqnarray}
   \begin{array}{lcl}
      \zeta_0    & = & \zeta_{00} \: + \: \zeta_{01} \, t \: + \:
                       \zeta_{02} \; t^2 \\
      \theta_0   & = & \theta_{00} \: + \: \theta_{01} \, t \: + \: 
                       \theta_{02} \, t^2 \\
       z         & = & z_{0\hphantom{0}} \: + \: z_{1\hphantom{0}} \, t \: + \: 
                       z_{2\hphantom{0}} \, t^2  \\
      \epsilon_0 & = & \epsilon_{00} \: + \: \epsilon_{01} \, t \: + \:
                       \epsilon_{02} \, t^2                         \\
      \Delta\psi & = & \displaystyle \sum_j^2
                       p_j \, \sin \ ( \alpha_{j} + \beta_{j} \, t ) \\
                       \hphantom{-}\hspace{-5em}\hphantom{-} \vspace{-3ex} \\
      \Delta\epsilon
                 & = & \displaystyle \sum_j^2 
                        e_j \, \cos \ ( \alpha_{j} + \beta_j \, t ) 
                       \hphantom{-}\hspace{-5em}\hphantom{-} \\
      S   & = & S_0 + \pi - E_0 \: + \: 
                (\Omega_n + \zeta_{01} + z_1 - E_1 ) \, t \: + \:
                    (\zeta_{02} + z_2 - E_2 ) \, t^2
                       \hphantom{-}\hspace{-5em}\hphantom{-} \\
          &   & + \;  \Delta\psi \, \cos\epsilon_0 \: - \:
                    \displaystyle
                    \sum_i^2 \left( E^c_i \cos \gamma_{i} \,t + 
                                    E^s_i \sin \gamma_{i} \, t \right) 
      \quad .
   \end{array}
\label{e:e9}\end{eqnarray}
}
  Here $t$ is TAI time elapsed from 2000 January 01, 12 hours. 
It should be noted that expressions \ref{e:e9} differs from those used in 
the framework of the traditional approach~\citep{r:iers2003}. Some of these 
parameters were taken from theory, some of them were found with the LSQ fit 
of time series of adjustments of pole coordinates and UT1--TAI. The numerical 
values of parameters used in data analysis are presented in the online 
Tab.~\ref{t:t1}. The rms of the adjustments of the perturbational vector
of the Earth rotation with respect to the a~priori matrix presented in 
expressions~\ref{e:e8} and~\ref{e:e9} over the period of 1984--2006 was 
less than $2.0 \cdot 10^{-6}$~rad for each component. 

\subsection{Basic estimation model}

   Several solutions were produced. Each solution used the basic 
parameterization which was common for all runs, and a specific parameterization
for an individual solution. Basic parameters belong to one of the 
three groups: 
\begin{itemize}
       \item [---]{\it global} (over the entire data set): positions of 
                          598 sources and proper motions of 169 sources;
                          positions and velocities of 44 stations; 
                          coefficients for the expansion into B-spline 
                          basis of positions of 7 stations, DSS15, DSS65,
                          GILCREEK, HRAS\_085, MEDICINA, MOJAVE12, PIETOWN,
                          \citep{r:nlm}; coefficients for harmonic position
                          variations of all sites at the annual and the
                          semi-annual frequency \citep{r:harpos}; coefficients
                          of the B-spline for modeling the perturbational 
                          Earth's rotation vector $\vec{q}_e(t)$ at a set of 
                          equidistant knots with a time span of 3 days 
                          for components~1 and 2 and with a time span 
                          of 1~day for component 3.

       \item [---]{\it local}  (over each session):  
                          tilts of the local symmetric axis of the atmosphere
                          for all stations and their rates, station-dependent 
                          clock functions modeled by second order polynomials, 
                          baseline-dependent clock offsets.

       \item [---]{\it segmented} (over 0.33--1.0 hours): coefficients of 
                          linear spline that models atmospheric path delay
                          (0.33 hour segment) and clock function 
                          (1 hour segment) for each station. The estimates 
                          of clock function absorb uncalibrated instrumental
                          delays in the data acquisition system.
\end{itemize}

  The rate of change for the atmospheric path delay and clock function between 
two adjacent segments was constrained to zero with weights reciprocal to 
$ 1.1 \cdot 10^{-14} $ and \mbox{$2\cdot10^{-14}$}, respectively, in order 
to stabilize solutions. Strong no-net-translation and no-net-rotation 
constraints were imposed on the adjustments of site positions and velocities,
as well as no-net-rotation constraints were imposed on adjustments of source 
positions and proper motions, in order to solve the LSQ problem of 
incomplete rank. Weak stabilizing constraints were imposed on B-spline 
coefficients to constrain to zero $\vec{q}_e(t)$, its first and second 
derivatives at each knot. The reciprocal weights of constraints for the 
values of $q_e(t)$, components 1 and 2 were \mbox{$5 \cdot 10^{-7}$ rad}; 
the reciprocal weights for the first time derivatives were 
\mbox{$5 \cdot 10^{-14}$ rad s${}^{-1}$}, and for the 2nd derivatives 
\mbox{$3 \cdot 10^{-19}$ rad s${}^{-2}$}. The reciprocal weights in 
$q_e(t)$, $\dot{q}_e(t)$, $\ddot{q}_e(t)$ for component 3 were
\mbox{$5 \cdot 10^{-7}$ rad, $3 \cdot 10^{-14}$ rad s${}^{-1}$}, and
\mbox{$6 \cdot 10^{-19}$ rad s${}^{-2}$}. 

\subsection{VLBI solutions}

  Several solutions have been computed with different a~priori vectors of 
the perturbational rotation $\vec{q}_a(t)$ and with a different set of 
harmonic constituents of the vector $\vec{q}_e(t)$.

  In solution~A, the a~priori vector $\vec{q}_a(t)$ was set to zero. 
A set of sine and cosine amplitudes of the perturbational rotation vector was
estimated as global parameters. The frequencies of these estimates were 
selected according to the following process. First, the frequencies for 
components 1 and 2 of this vector in the range of the near-diurnal 
retrograde wobble 
$[-7.310955 \cdot 10^{-5}, \: -7.298755 \cdot 10^{-5} \: \mbox{rad} \, 
\mbox{s}^{-1}]$ and hypothetical near-diurnal prograde wobble (see, for 
instance, \citet{r:Deh93}) 
$[-7.284405 \cdot 10^{-5}, \: -7.273275 \cdot 10^{-5} \: 
\mbox{rad} \, \mbox{s}^{-1}]$ 
were sampled with step $9.0 \cdot 10^{-9} \: \mbox{rad} \, \mbox{s}^{-1}$. 
This step of frequency sequences corresponds to $ 2\pi/\Delta T $ where 
$\Delta T$ is the interval of observations, 22.6 years. Then the frequencies 
of constituents with the amplitudes exceeding $ 1 \cdot 10^{-11} \: \mbox{rad}$
in the REN--2000 expansion were added to the list of constituents in order 
of decreasing their amplitudes, provided the minimal difference in the 
frequencies of the constituents was less than 
$9.0 \cdot 10^{-9} \: \mbox{rad} \, \mbox{s}^{-1}$. If two constituents have 
difference in frequencies less than that, the constituent with smaller 
amplitude was discarded.  

  Second, all constituents at positive and negative diurnal, semi-diurnal, 
and ter-diurnal bands of the tide-generating potential with amplitudes greater 
than 0.003 of the amplitude of the $ M_2 $ tide were selected. At negative 
frequencies, all three components of the vector $\vec{q}_e(t)$ were estimated, 
while only the components~1 and 2 were estimated at positive frequencies. 
The same rejection criteria for the constituents with close frequencies 
was enforced. 

  Third, the harmonic signal in components~1 and 2 of $\vec{q}_e$ at 
the Chandler and annual frequencies, and the harmonic signal in the 
component~3 at 14~frequencies of zonal tides with amplitudes greater 
than 3\% of the amplitude of the tide generating potential at the 
{\it Mf} frequency $5.323414 \cdot 10^{-6}$~rad~s${}^{-1}$ were estimated. 
Decorrelation constraints were imposed on estimates of B-spline coefficients.

  Solution~B is similar to solution~A, but constituents at the 76~frequencies
that have a close companion, a sidelobe, and that satisfy the criteria in 
Sect.~\ref{s:basis}, \enskip
$ A_2\frac{\displaystyle\strut \omega_2 - \omega_1}
          {\displaystyle\strut \Delta \omega_{min}} > 1 \cdot 10^{-11} 
           \: \mbox{rad}$, 
were not rejected as in solution~A, but remained on the list. Strong 
constraints with reciprocal weight $ 10^{-24} \: \mbox{rad}^2 $ in the form 
of Eq.~\ref{e:e7} were imposed. The purpose of this solution was 
to investigate the effects of omitted sidelobes.

  A family of solutions~C was computed. The purpose of this solution 
was to evaluate a harmonic signal at non-tidal frequencies. 
In addition to the frequencies used in the solution A, a possible non-tidal 
signal was sought in eight frequency bands at the frequencies equally 
sampled with the step $9.0 \cdot 10^{-9} \: \mbox{rad} \, \mbox{s}^{-1}$. 
The low and high edges of each frequency band are shown in Tab.~\ref{t:t2}. 
The total number of constituents of the perturbational rotation vector 
estimated for components~1 and 2 was 20\,093 and and for component 3 was 9885.
Since the total number of global parameters was at a level of 70\,000, 
well beyond the capabilities of a personal computer, 20 individual solutions 
were performed. In each individual solution of family~C, the sine and cosine 
amplitudes of constituents at $\sim\! 2000\:$ frequencies were estimated. 
The estimates of constituents are correlated at a level of 0.02--0.20, 
since other common parameters were estimated, such as source coordinates and 
station positions. Strictly speaking, the procedure of estimating 
the spectrum by parts is not perfectly correct. But it was assumed that 
such a  procedure may result in a false detection, but not miss the 
signal present in the data. In the final run of a solution of family~C, 
680 non-tidal frequencies were selected. The signal was detected at 
a 95\% confidence level at these frequencies in the previous runs. 

\begin{table}
  \caption{The range of frequency bands for estimation of a non-tidal
           signal. The last column refers to components of $\vec{q}_e(t)$
           vector of perturbational rotation that were estimated.}
  \begin{tabular}{l@{\quad}l@{\quad}l}
     \hline
     \multicolumn{2}{c}{Frequency band} & Components \\
     low & high                         &            \\
     \hline
         $ -2.95 \cdot 10^{-4} \:\: \mbox{rad} \, \mbox{s}^{-1} $  &
         $ -2.85 \cdot 10^{-4} \:\: \mbox{rad} \, \mbox{s}^{-1} $  &
         1, 2, 3  
\\
         $ -2.27 \cdot 10^{-4} \:\: \mbox{rad} \, \mbox{s}^{-1} $  &
         $ -2.07 \cdot 10^{-4} \:\: \mbox{rad} \, \mbox{s}^{-1} $  &
         1, 2, 3  
\\
         $ -1.60 \cdot 10^{-4} \:\: \mbox{rad} \, \mbox{s}^{-1} $  &
         $ -1.32 \cdot 10^{-4} \:\: \mbox{rad} \, \mbox{s}^{-1} $  &
         1, 2, 3  
\\
         $ -0.97 \cdot 10^{-4} \:\: \mbox{rad} \, \mbox{s}^{-1} $  &
         $ -0.52 \cdot 10^{-4} \:\: \mbox{rad} \, \mbox{s}^{-1} $  &
         1, 2, 3  
\\
         \hphantom{$-\!$}
         $  0.52 \cdot 10^{-4} \:\: \mbox{rad} \, \mbox{s}^{-1} $  &
         \hphantom{$-\!$}
         $  0.97 \cdot 10^{-4} \:\: \mbox{rad} \, \mbox{s}^{-1} $  &
         1, 2  
\\
         \hphantom{$-\!$}
         $  1.32 \cdot 10^{-4} \:\: \mbox{rad} \, \mbox{s}^{-1} $  &
         \hphantom{$-\!$}
         $  1.60 \cdot 10^{-4} \:\: \mbox{rad} \, \mbox{s}^{-1} $  &
         1, 2  
\\
         \hphantom{$-\!$}
         $  2.07 \cdot 10^{-4} \:\: \mbox{rad} \, \mbox{s}^{-1} $  &
         \hphantom{$-\!$}
         $  2.27 \cdot 10^{-4} \:\: \mbox{rad} \, \mbox{s}^{-1} $  &
         1, 2 
\\
         \hphantom{$-\!$}
         $  2.85 \cdot 10^{-4} \:\: \mbox{rad} \, \mbox{s}^{-1} $  &
         \hphantom{$-\!$}
         $  2.95 \cdot 10^{-4} \:\: \mbox{rad} \, \mbox{s}^{-1} $  &
         1, 2 
\\
     \hline
  \end{tabular}
  \label{t:t2}
\end{table}

  The purpose of solution D was to investigate whether the process of 
frequency selection for solution~A picked up all the signals in the diurnal 
band. In this solution the a~priori vector $\vec{q}_a(t)$ was generated from 
the REN--2000 nutation expansion.  It had all the constituents with 
amplitudes greater than $1 \cdot 10^{-9}$ rad, except the constituents with
frequencies \mbox{$ -7.331937 \cdot 10^{-5}$} 
            \mbox{$ -7.293186 \cdot 10^{-5}$},  
            \mbox{$ -7.291047 \cdot 10^{-5}$}, and
            \mbox{$ -7.252295 \cdot 10^{-5}$} rad~s${}^{-1}$, because they 
had already been included in the matrix $ \widehat{\mathstrut\cal M}_a(t)$. 
For selection of the frequencies, at which the harmonics of the 
perturbational rotation vector was to be estimated, the time series 
of $\vec{q}_e^B(t) + \vec{q}_a^B(t) - \vec{q}_a^C(t)$ on the interval 
[1984.0, 2006.6] with a step of 0.125 days were produced, and its power 
spectrum was computed. Here $\vec{q}_e^B(t)$ is the vector of the adjustments 
of perturbational rotation in solution B and vectors $\vec{q}_a^B$, where
$\vec{q}_a^D$ are the a~priori perturbational vectors used. 
The frequencies of constituents run from 0 through $ 2.9 \cdot 10^{-4} \; 
\mbox{rad} \, \mbox{s}^{-1} $ with step $ 9.0 \cdot 10^{-9} \; 
\mbox{rad} \, \mbox{s}^{-1}$. Not all of them were estimated, but only those 
that had the frequency by module less than $ 9.3 \cdot 10^{-5} \; 
\mbox{rad} \, \mbox{s}^{-1}$ and that satisfied one of the three conditions: 
a)~to be in the range $[7.20 \cdot 10^{-5}, \:\: 7.38 \cdot 10^{-5}] \; 
\mbox{rad} \, \mbox{s}^{-1}$ for taking the near-diurnal wobble 
and the signal excited by the atmosphere into account, or
b)~to have the square root of the power greater than $ 1 \cdot 10^{-11}$~rad 
for taking the signal at tidal frequencies into account, or 
or c)~to be in the range of non-tidal frequencies identified in solutions~C. 
In total, 3676 parameters at 1706 frequencies, which are a multiple of the 
frequency $ 9.0 \cdot 10^{-9} \: \mbox{rad} \, \mbox{s}^{-1}$, were selected. 

  The software Calc/Solve was used for these solutions. Results of analysis
are available on the Web at http://vlbi.gsfc.nasa.gov/erm.

\section{Discussion of results}
\label{s:res}

\subsection{Differences with respect to the traditional approach}

   Analysis of the results showed that the approach for the direct estimation 
of the perturbational rotation vector in the form of expansion into basic
functions from VLBI observations allows us to represent the Earth's rotation 
adequately. The fit of the least square solutions, 21.9~ps, was 
the same as the fit in solutions that followed the traditional time series 
approach. The slowly varying component was compared with the USNO EOP model. 
The USNO Finals EOP time series with 1~day steps was derived by averaging 
results of analysis of GPS and VLBI observations and smoothing. To compare 
results, the USNO series of polar motion, $X_p, Y_p$, and UT1--TAI with
respect to the a~priori Earth rotation model in expressions~\ref{e:e9}:
\begin{eqnarray}
   \begin{array}{lll}
      q^u_1 & = & Y_p(t) \vspace{0.5ex} \\
      q^u_2 & = & X_p(t) \vspace{0.5ex} \\
      q^u_3 & = & \kappa (UT1-TAI)(t) + (E_0 + E_1 \, t + E_2 \, t^2) \: - \\ 
                & & \displaystyle
                        \sum_i^2 \left( E^c_i \cos \gamma_i \,t + 
                         E^s_i \sin \gamma_i \, t \right) 
                         + \int\limits_{t_0}^{t} (\dot{\psi} + \Delta\dot{\psi}) 
                           \, \Delta\epsilon \sin \epsilon_0 \, dt 
                         \hphantom{-}\hspace{-3em}\hphantom{-}
   \end{array}
   \label{e:e10}
\end{eqnarray}
  where $\kappa = -(\Omega_n + z_{01} + \zeta_{01}) \cdot 86400/2\pi$, 
and parameters $ E_0, E_1,  E_2, E^c_i, E^s_i, \gamma_i, \psi, \Delta\psi,
\Delta\epsilon, \epsilon_0, \Omega_n, z, \zeta $ are from 
expressions~\ref{e:e9}. The coefficients of the interpolation spline 
for $\vec{q}^u$ were computed. These coefficients form the USNO Earth
rotation model as a continuous function of time. Since the GPS results 
almost entirely dominate components~1 and 2 of the Earth rotation 
vector from that model, they can be considered independent from our analysis 
of VLBI observations. 

\begin{figure}
   \raisebox{0.06\textwidth}{\includegraphics[width=0.13\textwidth,angle=90]{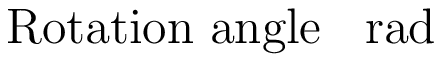}}
   \includegraphics[width=0.47\textwidth,clip]{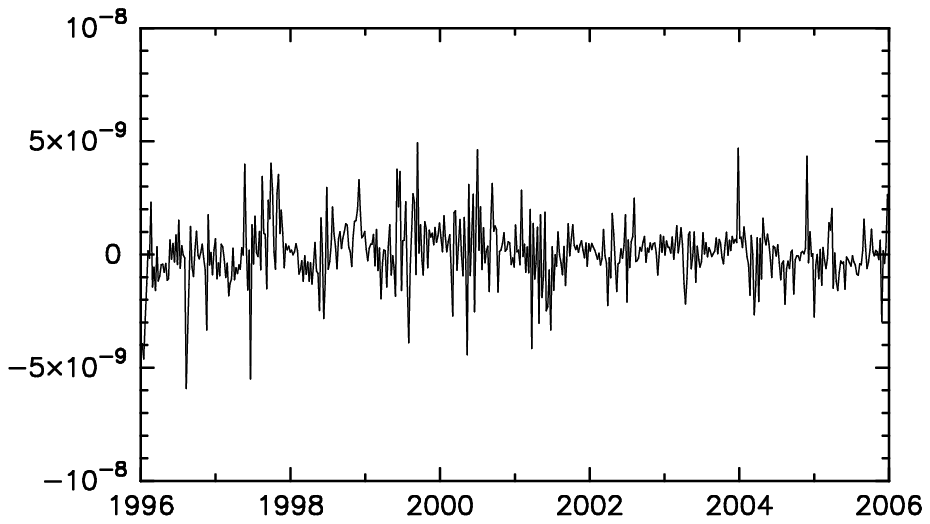}
   \par\vspace{-0.5ex}
   \parbox{0.49\textwidth}{\hfill Time in years}
   \caption{The component 1 of the residual perturbational vector with respect
            to the Earth rotation USNO Finals EOP model.}
   \label{f:erm_e1.eps}
\end{figure}

  The differences for component~1 between the USNO model and our results 
from solution~B after removal the contribution of harmonic variations with 
periods less than 2 days are shown in Fig.~\ref{f:erm_e1.eps}. No pattern 
of systematic differences is revealed. The statistics of these differences 
for all three components of the small vector of the Earth rotation and their 
time derivatives computed at the equidistant grid with time interval 2.5~hours 
are presented in the 1st row of Tab.~\ref{f:erm_dif.tab}.

  Since the VLBI observations are not carried out continuously due to budget
limitations, the accuracy of the Earth orientation model is the highest within
an interval of observations and the lowest at moments of time when there were
no observations. In the framework of the traditional approach, the EOP are 
estimated on moments of time in the middle of a 24 hour observing 
session. The statistics of the differences of the EOP series from analysis 
of VLBI observations gsf2006c\footnote{Available on the Web at
http://vlbi.gsfc.nasa.gov/solutions/2006c} for moments in the middle of
1426 observing sessions are shown in the 3rd row of Tab.~\ref{f:erm_dif.tab}. 
For comparison, the EOP were computed from results of solution~B at exactly 
the same epochs, and these statistics of the differences with respect to the 
USNO model are presented in the 2nd row of this table. 

\begin{table*}
   \caption{The weighted root mean squares of the differences between estimates
            of the Earth rotation model from analysis of VLBI observations 
            and the USNO Finals EOP model for the period of 
            [1996.0, 2006.0]. The statistics in rows 1 and 2 correspond to
            solution~B, which follows the proposed approach. The statistics 
            in row 3 correspond to solution gsf2006c, which follows 
            the traditional approach. }
   \begin{tabular}{l@{\qquad}ccc@{\qquad}ccc}
     \hline
     Solution  & $q_1$        &  $q_2$        &  $q_3$        &  
                 $\dot{q}_1$  &  $\dot{q}_2$  &  $\dot{q}_3$  
     \vphantom{$\bigl($} \\
     \hline
     ERM all   & $ 0.79 \cdot 10^{-9}  $ rad &
                 $ 0.99 \cdot 10^{-9}  $ rad &
                 $ 0.64 \cdot 10^{-9}  $ rad &
                 $ 0.78 \cdot 10^{-14} $ rad\,s${}^{-1}$ &
                 $ 1.16 \cdot 10^{-14} $ rad\,s${}^{-1}$ &
                 $ 0.92 \cdot 10^{-14} $ rad\,s${}^{-1}$ 
     \\
     ERM exp   & $ 0.58 \cdot 10^{-9}  $ rad &
                 $ 0.69 \cdot 10^{-9}  $ rad &
                 $ 0.52 \cdot 10^{-9}  $ rad &
                 $ 0.77 \cdot 10^{-14} $ rad\,s${}^{-1}$ &
                 $ 1.15 \cdot 10^{-14} $ rad\,s${}^{-1}$ &
                 $ 0.81 \cdot 10^{-14} $ rad\,s${}^{-1}$ 
     \\
     gsf2006c  & $ 0.55 \cdot 10^{-9}  $ rad &
                 $ 0.57 \cdot 10^{-9}  $ rad &
                 $ 0.42 \cdot 10^{-9}  $ rad &
                 $ 1.89 \cdot 10^{-14} $ rad\,s${}^{-1}$ &
                 $ 2.00 \cdot 10^{-14} $ rad\,s${}^{-1}$ &
                 $ 1.52 \cdot 10^{-14} $ rad\,s${}^{-1}$ 
     \\
     \hline
   \end{tabular}
   \label{f:erm_dif.tab}
\end{table*}

  Analysis of statistics shows that the differences in components~1 and 2
of the Earth's orientation according to the proposed and traditional approaches
do not exceed 20\%. At the same time the proposed approach gives the estimates 
of all the components of the Earth's angular velocity vector by a factor of 
1.5--2.0 closer to the GPS results than the estimates produced in the framework
of the traditional approach. According to the traditional approach, the EOP 
rates and nutation daily offsets are computed for each session independently, 
which makes them less stable. With the proposed approach, at a given epoch 
several experiments contribute to estimates of EOP rate, which makes them 
more robust.

\begin{figure}
   \raisebox{0.06\textwidth}{\includegraphics[width=0.13\textwidth,angle=90]{rotang.eps}}
   \includegraphics[width=0.47\textwidth,clip]{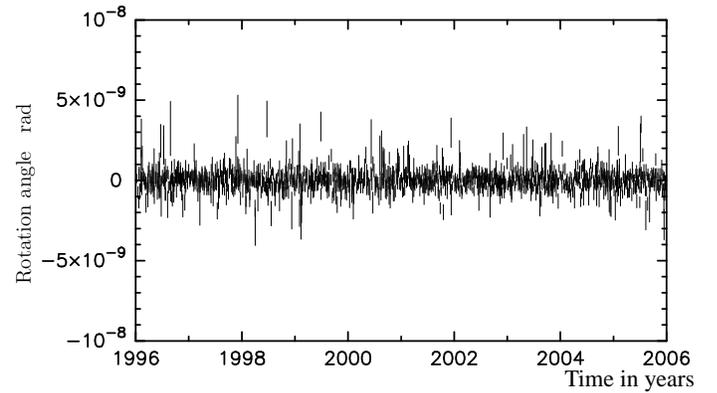}
   \par\vspace{-0.5ex}
   \parbox{0.49\textwidth}{\hfill Time in years}
   \caption{The time series of the estimates of the daily offsets of nutation
            in obliquity when the empirical Earth rotation model from 
            solution~B was used as the a~priori. The wrms is 
            $3.9 \cdot 10^{-10} \:$ rad.}
   \par\vspace{-2ex}
   \label{f:eps_erm}
\end{figure}

\begin{figure}
   \raisebox{0.06\textwidth}{\includegraphics[width=0.13\textwidth,angle=90]{rotang.eps}}
   \includegraphics[width=0.47\textwidth,clip]{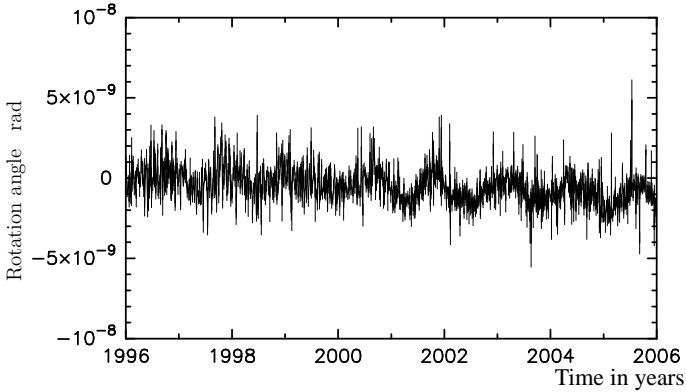}
   \par\vspace{-0.5ex}
   \parbox{0.49\textwidth}{\hfill Time in years}
   \caption{The time series of the estimates of the daily offsets of nutation
            in obliquity when the MHB2000 nutation expansion 
            was used as the a~priori. The wrms is 
            $9.8 \cdot 10^{-10} \:$ rad.}
   \label{f:eps_trad}
\end{figure}

  Analysis of the differences in amplitudes of the harmonic terms of 
components~1 and 2 of the vector of perturbational rotation at the retrograde 
diurnal band with respect to the semi-empirical MHB2000 expansion 
\citep{r:mhb2000}, showed they can reach 0.2~nrad for some terms. Detailed 
analysis of these differences is beyond the scope of the present paper. 
In order to test results, the empirical Earth rotation model from 
solution~B was used as the a~priori for the solution that estimated the 
time series of daily offset to nutations. The weighted root mean square of 
the differences for the period of [1996.0, 2006.0] is 0.39 nrad when results 
of solution~B were used as the a~priori, and 0.98 nrad when the MHB2000 was 
used. The daily offsets to nutation in obliquity $\Delta\epsilon(t)$ with 
respect to both models are shown in Figs.~\ref{f:eps_erm}--\ref{f:eps_trad}.

\subsection{Harmonic components in the Earth's rotation}

\begin{figure}
   \raisebox{0.06\textwidth}{\includegraphics[width=0.13\textwidth,angle=90]{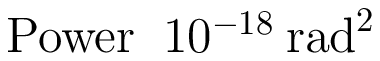}}
   \includegraphics[width=0.47\textwidth,clip]{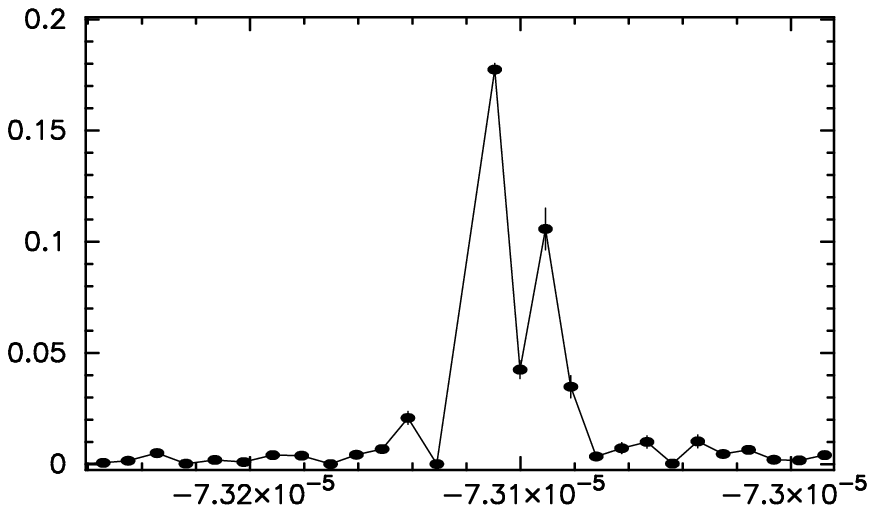}
   \par\vspace{-1ex}
   \parbox{0.49\textwidth}{\hfill rad s${}^{-1}$}
   \caption{The power spectrum of the estimates of the quasi-diurnal variations 
            of components~1 and 2 of the perturbational vector of the Earth's
            rotation from solution~A in the vicinity of the frequency of the 
            near-diurnal free wobble. The estimate for the frequency 
            \mbox{$-7.312026 \cdot 10^{-5} \: \mbox{rad} \: \mbox{s}^{-1}$}, 
            which corresponds to the tidal frequency $\psi_1$, is not shown.}
   \label{f:fcn}
\end{figure}

  Analysis of estimates of the harmonic components showed excessive power 
near the frequency of the near-diurnal retrograde wobble, as was expected. 
The spectrum turned out rather broad, spanning a rather wide band, and it
partly overlaps with the tidal frequency \mbox{$-7.312026 \cdot 10^{-5} \: 
\mbox{rad} \: \mbox{s}^{-1}$} that corresponds to the annual retrograde 
nutation as shown in Fig.~\ref{f:fcn}. It was found by \citet{r:Her86} 
that the near-diurnal wobble cannot be represented by a purely harmonic model 
with a constant amplitude. This means that when this component of the Earth's 
rotation is represented in the frequency domain, several constituents in the
spectrum will correspond to it. 

\begin{figure}
   \raisebox{0.06\textwidth}{\includegraphics[width=0.13\textwidth,angle=90]{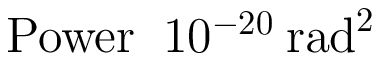}}
   \includegraphics[width=0.47\textwidth,clip]{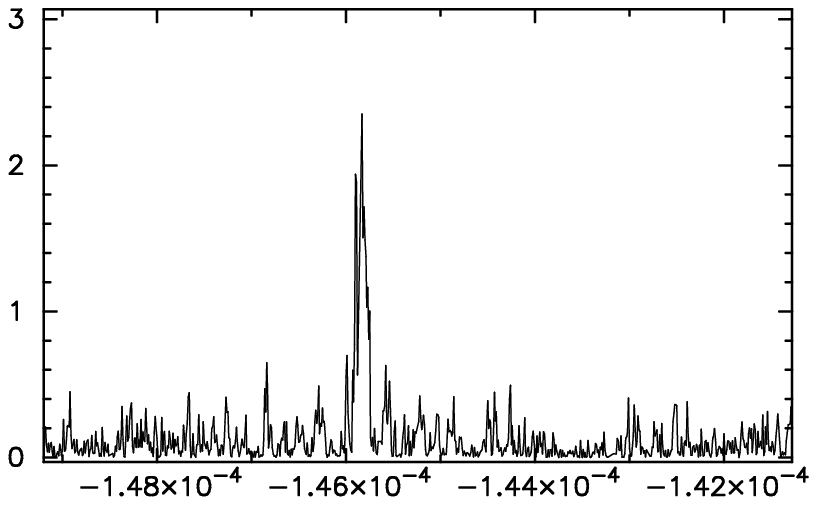}
   \par\vspace{-1ex}
   \parbox{0.49\textwidth}{\hfill rad s${}^{-1}$}
   \caption{The power spectrum of the estimates of the quasi-diurnal variations 
            of components~1 and 2 of the perturbational vector of the Earth's
            rotation from solution~C in the vicinity of the $-K_2$ frequency. 
            The estimate for the frequency 
            \mbox{$ -1.458423 \cdot 10^{-4} \: \mbox{rad} \: \mbox{s}^{-1}$}, 
            which corresponds to the $-K_2$, is not shown.}
   \label{f:k2}
\end{figure}

\begin{figure}
   \raisebox{0.06\textwidth}{\includegraphics[width=0.13\textwidth,angle=90]{power_k2.eps}}
   \includegraphics[width=0.47\textwidth,clip]{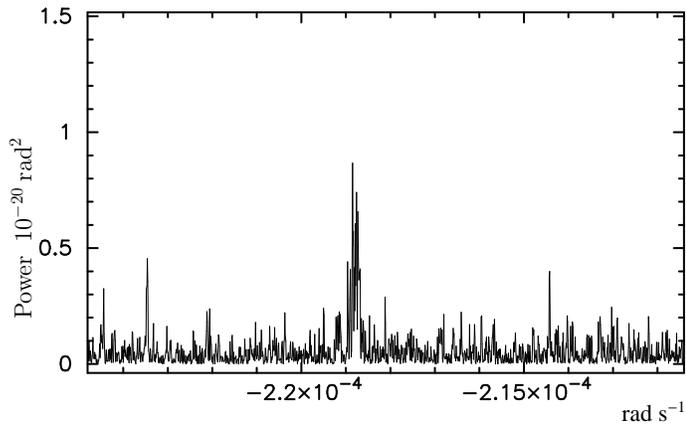}
   \par\vspace{-1ex}
   \parbox{0.49\textwidth}{\hfill rad s${}^{-1}$}
   \caption{The portion of the power spectrum of estimates of the ter-diurnal 
            variations of components~1 and 2 of the perturbational vector 
            of the Earth's rotation vector in the ter-diurnal band. The broad 
            peak is seen near the $-K_3$ frequency.}
   \label{f:k3}
\end{figure}

  Analysis of the results of the C~family solutions revealed several 
constituents with the non-tidal signal. The spectrum of components~1 and 2 
in the vicinity of the frequency $-2\Omega_n$, i.e. the $K_2$ tide, turned out 
rather broad. The excerpt of the power spectrum produced from estimates of 
sine and cosine amplitudes of the components~1 and 2 is shown in 
Fig.~\ref{f:k2}. This signal cannot be attributed to the spectral leakage, 
since no excessive power was found in the vicinity of even a stronger tide 
at the $M_2$ frequency. A relatively broad-band signal in the vicinity of
the $-3\Omega_n$ frequency, i.e. $K_3$, was found at the 3rd~component 
of the rotation vector. The excerpt of the power spectrum produced from 
estimates of sine and cosine amplitudes of the component~3 is shown in 
Fig.~\ref{f:k3}. A weaker signal in the estimates can also be revealed in the
vicinity of the $-4\Omega_n$ frequency. A similar signal can 
be seen at prograde frequencies in the vicinity $K_2, K_3, K_4$ at 
components~1 and 2\footnote{Since component~3 was considered as a real 
value process, its spectral power at negative frequencies is the same 
as at positive frequencies.}.

  Another peculiarity of the spectrum are sharp peaks at frequencies 
$ \pm 4/5\Omega_n,\, \pm 6/5\Omega_n$ at a level of 2--7$\sigma$ above the
noise level. No convincing explanation was found, but it is suspected 
that this signal in the estimates may be an artifact caused by errors in 
modeling by analogy with a detection of a very strong signal in estimates 
of the harmonic constituents of the perturbational Earth's rotation from 
GPS time series at frequencies that are multiple to the diurnal frequency: 
$S_1, S_2, S_3, S_4$, etc., reported by \citet{r:rotgps}. It should be noted 
that no non-tidal signal at $S_3, S_4$ frequencies is seen from analysis 
of VLBI group delays.

    Solution~D did not reveal other missed harmonic signals in the diurnal 
band. 

\subsection{Error analysis}

  The formal uncertainties of the amplitudes on harmonics constituents can be
evaluated on the basis of the signal-to-noise ratio of fringe phases 
used for computing group delay by invoking the law of error propagation.
These uncertainties are in a range of 5--12~prad. Analysis of the estimates 
of the constituents at the frequency bands where no tidal or no-tidal
signal was detected provides a more reliable measure of noise in adjustments.
It is 16~prad for components~1,2 and 13~prad for component~3 for the 
diurnal band; 13~prad and 10~prad for these components at other frequency 
bands. This corresponds to displacements of 0.06--0.12~mm at the Earth's 
surface. Evaluation of the level of systematic errors is more problematic. 
The major possible source of systematic errors is considered to be a residual 
motion of the individual stations. In fact, the rotation of the station 
polyhedron was evaluated, and it was assumed that the motion of this 
polyhedron is a representative measure of the Earth's rotation. This assumption 
is valid to the extent that residual horizontal motion of individual 
observing stations is negligible. \citet{r:harpos} estimated harmonic 
site position variations and found that the accuracy of modeling the horizontal 
motion of individual stations is at the level of 0.4~mm. In the case of the 
errors of modeling being completely uncorrelated, this error will be diluted as 
$ \sqrt{N_{eff}}$, where $\sqrt{N_{eff}} $ is the effective number of 
observing stations, 10--44, depending on how to define the effective number 
of stations. Unfortunately, the distribution of residual motions of stations
at tidal frequencies shows a pattern of a systematic behavior, which does 
not support the hypothesis of uncorrelated errors. A conservative estimate 
of the possible contribution of the unmodeled residual motion of the network 
of stations to the estimates of harmonic constituents of the perturbational 
rotation suggests a dilution factor of 2, i.e. the surface displacements 
$\sim\! 0.2$~mm. That means systematic errors may be two times greater 
than random errors.

  \citet{r:Deh03} investigated the influence of systematic errors due to 
the neglect of the modeling source structure. It was suggested to split 
the observed radio sources into two classes, ``stable'' and ``unstable'', 
and either to remove unstable sources from analysis or to estimate the time 
series of their positions. In this paper a different approach was used: 
proper motion of those sources that had a long enough history of observations 
was estimated. This method is supposed to reduce systemic errors in the 
estimates of the harmonic constituents in the perturbational rotation vector.

\section{Conclusion}
\label{s:conclude}

  It was demonstrated that the empirical Earth rotation model can be determined 
directly from observations over a period of 22 years using the least square 
estimation technique. The advantage of the proposed approach is that 
a continuous function describing the Earth's orientation is determined in one 
step without producing intermediate time series. The consistency between 
station positions, source coordinates, and the empirical Earth rotation model
is automatically achieved. Another advantage of the proposed approach 
is that a simplified a~priori model with only 31 numerical parameters 
is sufficient, while the traditional approach needs a complicated a~priori 
model of precession, nutation, high frequency harmonic variations of the 
Earth's rotation, and a filtered and smoothed time series of the Earth 
orientation parameters produced in the previous analysis, in total 
$46\,000$ numerical parameters \citep{r:iers2003}. 

  The traditional approach to describing the Earth's rotation follows the 
formalism of either Newcomb and Andoyer or \citet{r:Guinot79} and 
\cite{r:Cap86}, and involves such notions as the celestial intermediate pole, 
the point of the vernal equinox, the non-rotating origin, the ecliptic, and 
other axes, points, planes, and circles on the celestial sphere. The advantage 
of the empirical Earth rotation model is that it is conceptually simpler, 
since it is built entirely \emph{kinematically} and does not require 
introduction of intermediate points, axes, planes that are not observable.

  It was demonstrated that the empirical Earth rotation model derived from 
analysis of VLBI observations gives the differences with respect to the EOP 
derived from analysis of independent GPS observations at moments of observation
at the same level, within 20\%, as the differences of the VLBI EOP series 
produced with the traditional approach. The advantage of the proposed approach
is that the estimates of the EOP rates are a factor of 1.5--2.0 closer to
the GPS time series than the VLBI EOP rates estimated following the traditional 
approach. 

  When results of analysis of observations are compared with theoretical
predictions, two approaches can be taken: a)~the parameters that describe
empirical data are formulated through parameters of the theoretical models; 
b)~theoretical predictions are transformed to a form that can be unambiguously 
determined from the observations. Representation of the Earth's
rotation in the form of the expansion into basis functions establishes 
a foundation for the second approach.

  Scientific interpretation of the results of estimation of the
empirical Earth rotation model will be given in the next paper.

\acknowledgements

  This work was done while the author worked at the National Observatory of 
Japan, Mizusawa, as a visiting scientist. The author would like to 
thank S.~Manabe, T.~Sato, Y.~Tamura, D.~Rowlands, Ch.~Bizouard and 
an anonymous referee for useful discussion that helped to improve 
the manuscript.

\Online
\begin{table*}
  \caption{Numerical values of the a~priori Earth rotation model parameters 
           used in data reduction.}
  \begin{tabular}{llll}
      \hline
      Var  &  \multicolumn{1}{c}{Value}  &  Units  &  Source                \\
      \hline
      $ \zeta_{00}    $  &  $ \hphantom{-}1.140216587056520 \cdot 10^{-10}  $ & 
                      $  \mbox{rad}                  $ & \citet{r:Simon94}  \\
      $ \zeta_{01}    $  &  $ \hphantom{-}3.542805701761733 \cdot 10^{-12}  $ & 
                      $  \mbox{rad} \: \mbox{s}{^-1} $ & \citet{r:Simon94}  \\
      $ \zeta_{02}    $  &  $ \hphantom{-}1.471291601425477 \cdot 10^{-25}  $ & 
                      $  \mbox{rad} \: \mbox{s}^{-2} $ & \citet{r:Simon94}  \\
      $ \theta_{00}   $  &  $ \hphantom{-}9.909515599113584 \cdot 10^{-11}  $ & 
                      $  \mbox{rad}                  $ & \citet{r:Simon94}  \\
      $ \theta_{01}   $  &  $ \hphantom{-}3.079019263961936 \cdot 10^{-12}  $ & 
                      $  \mbox{rad} \: \mbox{s}^{-1} $ & \citet{r:Simon94}  \\
      $ \theta_{02}   $  &  $ -2.076601527511399 \cdot 10^{-25} $ & 
                      $  \mbox{rad} \: \mbox{s}^{-2} $ & \citet{r:Simon94}  \\
      $ z_{0}         $  &  $ \hphantom{-}1.140216587060519 \cdot 10^{-10}  $ & 
                      $  \mbox{rad} $  & \citet{r:Simon94} \\
      $ z_{1}         $  &  $ \hphantom{-}3.542805701761733 \cdot 10^{-12}  $ & 
                      $  \mbox{rad} \: \mbox{s}^{-1} $ & \citet{r:Simon94}  \\
      $ z_{2}         $  &  $ \hphantom{-}5.331975251279779 \cdot 10^{-25}  $ & 
                      $  \mbox{rad} \: \mbox{s}^{-2} $ & \citet{r:Simon94}  \\
      $ \epsilon_{00} $  &  $ \hphantom{-}0.409092629687089                 $ & 
                      $  \mbox{rad} $ & \citet{r:Simon94} \\
      $ \epsilon_{01} $  &  $ -7.191223191481661 \cdot 10^{-14}             $ & 
                      $  \mbox{rad} \: \mbox{s}^{-1} $ & \citet{r:Simon94}  \\
      $ \epsilon_{02} $  &  $ -7.399638794037328 \cdot 10^{-29} $ & 
                      $  \mbox{rad} \: \mbox{s}^{-2} $  & \citet{r:Simon94} \\
      $ S_0           $  &  $ \hphantom{-}1.753368559233960 $                & 
                      $  \mbox{rad}                  $  & \citet{r:Aoki82}  \\
      $ \Omega_n      $  &  $ \hphantom{-}7.292115146706979 \cdot 10^{-5} $ & 
                      $  \mbox{rad} \: \mbox{s}^{-1} $  & \citet{r:Aoki82} \\
      $ p_1           $  &  $ -8.377867467753367 \cdot 10^{-5}  $ & rad & \citet{r:ren2000a} \\
      $ p_2           $  &  $ -6.193374542381407 \cdot 10^{-6}  $ & rad & \citet{r:ren2000a} \\
      $ e_1           $  &  $ \hphantom{-}4.473817016047498 \cdot 10^{-5}  $ & rad & \citet{r:ren2000a} \\
      $ e_2           $  &  $ \hphantom{-}2.682642812740089 \cdot 10^{-6}  $ & rad & \citet{r:ren2000a} \\
      $ \alpha_1      $  &  $ \hphantom{-}2.182438855728973 $ & rad & \citet{r:Simon94} \\
      $ \alpha_2      $  &  $ \hphantom{-}3.506953516079786 $ & rad & \citet{r:Simon94} \\
      $ \beta_1       $  &  $ -1.069696206302000 \cdot 10^{-8} $ & 
                      $  \mbox{rad} \: \mbox{s}^{-1} $ & \citet{r:Simon94} \\
      $ \beta_2       $  &  $ \hphantom{-}3.982127698995000 \cdot 10^{-7} $ & 
                      $  \mbox{rad} \: \mbox{s}^{-1} $ & \citet{r:Simon94} \\
      $ E_0           $  &  $ \hphantom{-}2.260937669429621 \cdot 10^{-3}   $ & rad & LSQ fit \\
      $ E_1           $  &  $ \hphantom{-}1.029854567486117 \cdot 10^{-12}  $ & 
                      $  \mbox{rad} \: \mbox{s}^{-1}  $ & LSQ fit \\
      $ E_2           $  &  $ -7.875297448491237 \cdot 10^{-22}  $ & 
                      $  \mbox{rad} \: \mbox{s}^{-2} $   & LSQ fit  \\
      $ E^c_1         $  &  $ \hphantom{-}9.776692309499138 \cdot 10^{-5}  $ & 
                      $  \mbox{rad} $  & \citet{r:Dickman1993}  \\ 
      $ E^s_1         $  &  $ -6.857935725000193 \cdot 10^{-6}             $ & 
                      $  \mbox{rad} $  & \citet{r:Dickman1993}  \\ 
      $ E^c_2         $  &  $ \hphantom{-}3.783804480256964 \cdot 10^{-6}  $ & 
                      $  \mbox{rad} $  & LSQ fit \\
      $ E^s_2         $  &  $ \hphantom{-}2.878954568890594 \cdot 10^{-6}  $ & 
                      $  \mbox{rad} $  & LSQ fit \\
      $ \gamma_1      $  &  $ -1.069696206302000 \cdot 10^{-8} $ & 
                      $  \mbox{rad} \: \mbox{s}^{-1} $ & \citet{r:Dickman1993} \\
      $ \gamma_2      $  &  $ -1.183000000000000 \cdot 10^{-8} $ & 
                      $  \mbox{rad} \: \mbox{s}^{-1} $ & LSQ fit  \\
     \hline
  \end{tabular}
  \label{t:t1}
\end{table*}

\end{document}